\documentclass[aps,prd,twocolumn,groupedaddress,showpacs]{revtex4}

\usepackage{graphicx}%

\begin{document}

\title{The Inflationary Energy Scale in Braneworld Cosmology}  
\author{Rachael M. Hawkins and James E. Lidsey}
\affiliation{Astronomy Unit, School of Mathematical Sciences,
Queen Mary, University of London, Mile End 
Road, LONDON, E1 4NS, U.K.}

\begin{abstract}
Upper bounds on the energy scale at the end of inflation 
in the Randall--Sundrum type II braneworld scenario are 
derived. The analysis is made exact 
by introducing new parameters that represent extensions 
of the Hubble flow parameters. 
Only very weak assumptions about the form of the inflaton potential
are made. In the high energy and slow roll regime 
the bounds depend on the amplitude of 
gravitational waves produced during inflation and 
become stronger as this amplitude increases. 
\end{abstract}


\maketitle

\section{Introduction}

Inflation is presently the most favoured model for 
describing the earliest stages of the universe's history
\cite{simplest}. (For a review, see, e.g., Ref.
\cite{lidlyth}).
As well as resolving the horizon, flatness and monopole 
problems of the standard, big 
bang cosmology, it provides a quantum mechanical origin 
for generating the primordial density perturbations that subsequently 
formed large--scale structure through 
gravitational instability \cite{perturbations}.
The inflationary paradigm is strongly 
supported by recent observations from the 
Wilkinson Microwave Anisotropy Probe (WMAP) 
\cite{temp,wmap,wmap1,kogut,komatsu}. 
These indicate that the cosmic microwave background (CMB) power spectrum 
is consistent with the gaussian, adiabatic and nearly scale--invariant 
form expected from an inflationary epoch and 
that the universe is spatially flat to within the 
limits of observational accuracy \cite{wmap}. Moreover, 
the observed anti--correlation on degree angular scales 
between the temperature and polarization E--mode maps 
of the CMB \cite{kogut} 
provides strong evidence 
for the existence of correlations on length scales beyond the Hubble 
radius at the epoch of decoupling \cite{white}.

In the simplest class of inflationary models, 
the energy density of the universe is 
dominated by the potential energy of a single, scalar `inflaton' field, 
$\phi$, 
that slowly rolls down its self--interaction potential 
\cite{simplest}. 
In light of the above developments, there is a pressing need 
to understand the origin of the inflaton field within the 
context of a unified field theory.
Recent developments in our understanding of the 
non--perturbative properties of string/M--theory \cite{M,hw}
have led to the proposal that our observable, 
four--dimensional universe may be viewed as a domain wall 
embedded in a higher--dimensional `bulk' 
space \cite{standardbranerefs,RSII}. The standard model gauge interactions 
are confined to the four--dimensional hypersurface,
but gravitational interactions may propagate in the 
bulk dimensions. From a cosmological point of view, the Randall--Sundrum 
type II (RSII) scenario has generated considerable interest \cite{RSII}. 
In the simplest version 
of this scenario, 
our observable universe is represented by a codimension one 
brane embedded in five--dimensional, Anti--de Sitter (AdS) space. 
Although the fifth dimension is infinite in extent, corrections to 
Newton gravity remain undetectable if the warping of the non--factorizable 
geometry is sufficiently strong \cite{RSII}. 
(For a recent review of the cosmological implications 
of the RSII scenario, see, e.g., Ref. \cite{lidsey}). 

In the case where a single inflaton 
field is confined to the brane, 
the effective Friedmann equation is derived by 
projecting the five--dimensional Einstein field equations onto
a four--dimensional hypersurface sourced by the scalar field 
\cite{shiromizu}. The Friedmann equation acquires a quadratic 
dependence on the energy density that becomes important at 
high energy scales \cite{4a,bine,flanagan,shiromizu,othermoves}. 
Such a modification to the braneworld Friedmann equation
alters the dynamics of the inflaton field, compared with standard cosmology,
by providing an additional source of friction as the field 
rolls down its potential \cite{maartens}. 
This has made the RSII braneworld 
scenario attractive to inflationary model builders as it 
increases the number of 
inflationary 
potentials that can be considered. In the steep inflationary 
scenario, for example, 
potentials that can not 
support inflationary expansion in the standard cosmology
can provide a sufficient number of e--foldings of inflation in the 
braneworld scenario \cite{steep,hueylid1,guen}. 

In this paper, we focus
on the energy scale at the end of inflation in the RSII scenario. 
The inflationary energy scale is important for a number 
of reasons. Firstly, it is directly related to  
the reheating temperature of the universe immediately after inflation. 
If this temperature is too high, the overproduction of gravitinos and 
other moduli fields may violate constraints imposed at the epoch of 
nucleosynthesis \cite{susy}. 
Within the context of steep inflation, it is possible that reheating may 
proceed through gravitational particle production \cite{gpp}. 
However, in this case
the universe becomes dominated 
by gravitational waves before the epoch of nucleosynthesis if the energy 
scale at the end of inflation is lower than a critical value 
\cite{sami}. The inflationary energy scale is also 
important to 
quintessential inflationary models 
\cite{quintessence,hueylid1,sami,otherq,kostas}, 
where 
the inflaton field survives the reheating process to 
act today as the quintessence field. 
These models must satisfy the stringent coincidence constraint 
that the densities of dark energy and 
matter are comparable at the present epoch. Since the conditions at 
the end of inflation determine the subsequent evolution of the 
inflaton field, it is important to derive constraints on this earlier epoch. 
Moreover, one of the principle uncertainties in single field inflationary 
models at present is the number of e--foldings of inflationary 
expansion that elapsed 
between the epoch when observable scales first crossed the Hubble radius and 
the end of inflation \cite{dod,lid}. Since this quantity is related to the 
reheating process, and therefore the energy scale of inflation, 
constraints on such a scale can in principle lead 
to constraints on the number of e--foldings. 

Unfortunately the process which ends inflation is not directly 
observable, although the possible 
overproduction and subsequent evaporation of 
primordial black holes (PBHs) place strong constraints on 
the processes immediately after inflation \cite{pbh}. 
The earliest observational constraints available to us correspond to 
approximately 60 e-foldings 
before the end of inflation and arise from the CMB power spectrum at 
low multipoles. 
In particular, the amplitude of gravitational waves is directly related to 
the energy scale at this epoch \cite{gw}.  
In this paper, we relate 
the energy scale at the end of RSII inflation 
to the energy scale $N$ e--foldings 
before the end of inflation and therefore obtain upper 
bounds to the inflationary energy scale. This correspondence is made 
possible by generalizing an argument due to 
Lyth \cite{lyth} 
and Liddle \cite{liddle}
and employing an extension of the Hamilton--Jacobi 
formalism  
for scalar field dynamics in the early universe 
\cite{hj,lidseyplb,hawlid}.  
In particular, we relate the energy scales at the two epochs 
without any direct reference to the slow--roll approximation. 
Then, by assuming 
the slow--roll approximation to be valid 
$N$ e--foldings before the end of inflation,
we relate observable quantities, such as the 
scale dependence of the density perturbations 
and the relative amplitudes 
of the density and gravitational wave fluctuations, 
directly to the energy scale at the end of inflation 
in a way not previously demonstrated. 
The specific functional form of the inflaton potential need not be 
known and only  
weak and physically well--motivated assumptions are made. 

\section{Inflation in the RSII Braneworld Cosmology}

\subsection{Scalar Field Dynamics}

We begin by 
reviewing the formalism for rewriting 
the RSII braneworld scenario in a parallel fashion to that of the 
Hamilton--Jacobi framework of the standard scenario \cite{hawlid}. 

The Friedmann equation for the RSII scenario 
is 
given by 
\begin{equation}
\label{Friedmann}
H^2 = \frac{8\pi}{3 m^2_4} \rho \left(
1 + \frac{\rho}{2 \lambda} \right)  , 
\end{equation}
where $H \equiv \dot{a}/a$ represents the Hubble parameter, 
a dot denotes differentiation with respect to time, 
$\rho$ is the 
energy density of matter confined on the 
brane, $m_4^2$ is the four--dimensional
Planck mass, $\lambda$ is the brane tension and we have 
assumed that the four--dimensional 
cosmological constant is zero. 
The standard cosmic dynamics of Einstein gravity 
is formally recovered in the limit  
$\lambda \rightarrow \infty$, or 
equivalently, in the low--energy regime 
corresponding to $\rho \ll \lambda$. 

We assume throughout that the energy--momentum of the 
brane matter is dominated by a single inflaton 
field such that $\rho = \dot{\phi^2}/2 + V(\phi)$, 
where $V(\phi )$ represents the interaction potential. 
The equation of motion of the scalar field is then given by 
\begin{equation}
\label{eom}
\ddot{\phi}+3H \dot{\phi} + V'(\phi) =0   , 
\end{equation}
where the prime denotes differentiation with respect
to the field. 
The dynamics of each
particular inflationary 
model are determined by the Friedmann equation (\ref{Friedmann}) 
and the scalar field equation of motion (\ref{eom}) 
once the functional form of the inflaton potential has been specified. 

Algorithms for finding exact solutions to this 
system were developed in Ref. \cite{hawlid}
by defining a new dimensionless parameter, $y$: 
\begin{equation}
\label{ydefine}
\rho \equiv \frac{2\lambda y^2}{1-y^2}  , 
\end{equation}
in terms of the energy density, where
the restriction $y^2<1$ ensures the 
weak energy condition is satisfied. 
Substituting 
Eq. (\ref{ydefine}) into Eq. (\ref{Friedmann}) then 
yields the Hubble parameter as a function of $y$:
\begin{equation}
\label{yHubble}
H (y ) = \left( \frac{16\pi \lambda}{3m^2_4} 
\right)^{1/2} \frac{y}{1-y^2}  . 
\end{equation}
The Hubble parameter increases monotonically with 
increasing $y$ in the range of physical interest. 

It is reasonable to assume that the scalar field monotonically 
rolls down its potential during the final 
stages of inflation $(\dot{\phi} \ne 0)$. This implies that 
the inflaton field may be viewed as the dynamical variable and 
Eq. (\ref{eom}) may then be simplified to 
\begin{equation}
\label{first}
\rho' =-3H\dot{\phi}   .
\end{equation}
Substituting Eq. (\ref{ydefine}) 
into Eq. (\ref{first}) 
then yields
\begin{equation}
\label{yprime}
\dot{\phi} = - \left( \frac{\lambda m^2_4}{3\pi} \right)^{1/2}
\frac{y'}{1-y^2}  
\end{equation}
and it follows from Eqs. (\ref{yHubble}) and (\ref{yprime}) 
that the scale factor, $a(\phi)$, satisfies \cite{hawlid}
\begin{equation}
\label{ysatisfy}
y'a' =-\frac{4\pi}{m^2_4} ya   .
\end{equation}
Integration of Eq. (\ref{ysatisfy})
yields the scale factor 
in terms of a 
single quadrature with respect to the inflaton field:
\begin{equation}
\label{yscale}
a(\phi ) = \exp \left[ -\frac{4 \pi}{m^2_4} 
\int^{\phi} d \phi \frac{y}{y'}
\right]  
\end{equation}
and the inflaton potential is determined by substituting 
Eqs. (\ref{yHubble}) and (\ref{yprime}) into 
Eq. (\ref{Friedmann})
\cite{hawlid}:
\begin{equation}
\label{ypotential}
V (\phi ) = \frac{2\lambda y^2}{1 -y^2}  
- \frac{\lambda m^2_4}{6 \pi} \left( \frac{y'}{1-y^2} 
\right)^2   . 
\end{equation}

Thus, in the braneworld scenario
the function $y (\phi) $, as defined in Eq. (\ref{ydefine}),  
plays a similar role to that of the Hubble parameter, $H(\phi)$, 
of conventional scalar field cosmology. 
Employing 
the scalar field as the dynamical variable implies that the 
second--order system of 
equations (\ref{Friedmann}) and (\ref{eom}) 
can be reduced to the non--linear,  
first--order system (\ref{yprime}) and (\ref{ysatisfy}).
Since the Hubble parameter decreases monotonically 
with decreasing $y$, the problem of maximizing 
the energy scale at the end of inflation is transformed to that 
of maximizing $y$. 

\subsection{Slow--Roll and Hubble Flow Parameters}

We now proceed to parametrize the 
classical scalar field dynamics in the RSII scenario 
in terms of the parameter $y(\phi)$ and its derivatives.
In conventional inflationary models, the 
dynamics may be described by the `Hubble flow' parameters,
defined such that 
\begin{eqnarray}
\label{epsilon}
\epsilon_H (\phi) \equiv   - \frac{\dot H}{H^2} =  
\frac{m_{4}^{2}}{4\pi}\frac{H'^2}{H^2}  ,  \\
\label{eta}
\eta_H (\phi) = \frac{m_{4}^{2}}{4\pi} \frac{H''}{H}
\end{eqnarray}
and inflation proceeds for 
$\epsilon_H <1$.

An important parameter pair in the RSII scenario are the 
braneworld slow--roll 
parameters. These are defined by 
$\epsilon_B \equiv -\dot{H}/H^2$ and $\eta_B \equiv V''/(3H^2)$ 
and are given by \cite{maartens}
\begin{eqnarray}
\label{epsilonb}
\epsilon_B  \simeq \frac{m_4^2}{4 \pi} \, 
\left( \frac{V'}{V} \right)^2 \,
\left[ \frac{1+V/\lambda}{\left(2 + V/\lambda \right)^2} \right] \\
\label{etab}
\eta_B \simeq \frac{m_4^2}{8\pi} \left( \frac{V''}{V} \right) 
\left[ \frac{2\lambda}{2\lambda +V} \right]
\end{eqnarray}
in the slow--roll limit, 
$\dot{\phi}^2 \ll V$ 
and $|\ddot{\phi}| \ll H |\dot{\phi}|$.
Self--consistency 
of the slow--roll approximation requires that $ {\rm max} 
\{ \epsilon_B , | \eta_B | \} \ll 1$. These expressions are important when 
relating observable quantities such as the spectral indices and 
gravitational wave amplitudes directly to the inflaton potential and its 
first two derivatives, but only approximately describe the classical 
inflaton dynamics. Thus, in the low--energy 
limit, Eqs. (\ref{epsilonb}) and (\ref{etab}) reduce to the 
potential slow--roll parameters, $\epsilon_V \approx 
m_4^2V'^2/(16\pi V^2)$ and 
$\eta_V = m_4^2V''/(8\pi V)$, respectively, and not the Hubble flow 
parameters (\ref{epsilon}) and (\ref{eta}). 

In the previous Subsection it was shown that the 
variable $y(\phi)$ acts in an
analogous way to the Hamilton--Jacobi function $H(\phi)$. 
Motivated by these considerations, we 
now introduce two new parameters, 
$\beta(y)$ and $\gamma(y)$, defined such that \footnote{Since 
we are assuming 
throughout that 
the inflaton field is a monotonically varying function of time, 
we may specify $\dot{\phi} > 0$ without loss of generality. Eq. 
(\ref{yprime}) then implies that $y' <0$ and hence Eq. (\ref{betadefn}) 
implies that $\sqrt{\beta} = -\sqrt{m_4^2/4\pi} y'/y$.} 
\begin{eqnarray}
\label{betadefn}
\beta \equiv \frac{4\pi}{m_{4}^{2}} \frac{ \dot\phi^{2}}{H^{2}}  = 
\frac{m_{4}^{2}}{4\pi} \frac{y'^2}{y^2} \\
\label{gammadefn}
\gamma \equiv \frac{m_{4}^{2}}{4\pi} \frac{y''}{y}  \\
\label{betagamma}
\gamma = \beta - \sqrt{\frac{m_{4}^{2}}{16\pi}} \frac{\beta'}{\sqrt{\beta}}  .
\end{eqnarray}
Since $H \propto y$ in the low--energy regime $(\rho/2\lambda \rightarrow 0 , 
y \rightarrow 0$), it follows immediately that  
Eqs. (\ref{betadefn}) and (\ref{gammadefn}) reduce to 
Eqs. (\ref{epsilon}) and (\ref{eta}) in this limit. 
These parameters may therefore be 
viewed as generalizations to the RSII inflationary scenario
of the Hubble flow 
parameters of standard inflation. 
We refer to the parameters (\ref{betadefn}) and (\ref{gammadefn})
as the {\em braneworld flow parameters}. 
In principle an infinite hierarchy of such parameters
involving higher derivatives of $y(\phi )$ may be defined  
along the lines of Ref. \cite{infinite,infrev1}. 
We emphasize that no reference 
has yet been made to the slow--roll approximation. 
Indeed, it can be shown by combining Eqs. (\ref{yHubble}), 
(\ref{yprime}) and (\ref{betadefn}) that 
\begin{equation}
\label{epsilonbeta}
\beta = \left(\frac{1-y^2}{1+y^2} \right) \epsilon  ,
\end{equation}
where $\epsilon = -\dot{H}/H^2$. 

The number of e--foldings of inflationary expansion, 
$N \equiv \ln (a_{end} /a_N )= \int^{t_{end}}_{t_N} dt \, H$, 
that occur when the scalar field rolls from some 
value, $\phi_N$, to the value, $\phi_{end}$, 
is given exactly in terms of a quadrature involving $\beta (\phi)$:
\begin{equation}
\label{yefolds}
N =- \frac{4\pi}{m_{4}^{2}}\int^{\phi_{end}}_{\phi_{N}} 
d\phi\frac{y}{y'} 
= \sqrt{\frac{4\pi}{m_{4}^{2}}}\int^{\phi_{end}}_{\phi_{N}} 
\frac{d\phi}{\sqrt{\beta (\phi )}}   ,
\end{equation}
where the subscript `$N$' denotes the value $N$ e--foldings 
before the end of inflation and a subscript `end' denotes values 
at the end of inflation, respectively. 
Moreover, rearranging Eq. (\ref{betadefn}) and integrating yields a 
second quadrature in terms of $\beta$: 
\begin{equation}
\label{yendratio}
\frac{y_{end}}{y_{N}} = \exp \left[ 
-\sqrt{\frac{4\pi}{m_{4}^{2}}}\int^{\phi_{end}}_{\phi_{N}} 
d\phi \sqrt{\beta(\phi)}\right]  .
\end{equation}

The RSII braneworld cosmology 
may therefore be parametrized in terms of the 
three functions $y(\phi)$, 
$\beta[y (\phi )]$ and 
$\gamma [y (\phi)]$, which are analogous to the familiar functions 
arising in the Hamilton--Jacobi formalism of 
standard cosmological models:  $H(\phi)$, $\epsilon_H [H (\phi)]$ and 
$\eta_H [H (\phi)]$ \cite{hj,lidseyplb,hawlid}. 
The flow parameters (\ref{betadefn}) and 
(\ref{gammadefn}) play an important role in developing 
upper bounds on the inflationary energy scale
as they provide a relationship between the variable $y$ and its derivatives
without specifying the functional form of the 
inflaton potential or restricting the analysis to the slow--roll regime. 
If the functional form of $\beta (\phi)$ 
is known, Eqs. (\ref{yefolds}) 
and (\ref{yendratio}) may be solved to determine $y(\phi_{end})$. 
The magnitude of the Hubble parameter at this epoch then follows
directly from Eq. (\ref{yHubble}). 
We proceed in the following Section to derive an upper limit
on this parameter. 

\section{Upper Limits on the Braneworld Inflation Energy Scale}

We begin by noting that 
Eq. (\ref{yprime}) implies that 
$y$ is a monotonically decreasing function of time $(\dot{y} <0)$.
It then follows from Eq. 
(\ref{ydefine}) that the maximal value of the energy scale at the end 
of inflation is deduced by maximizing the value of $y_{end}$. 
Since 
inflation ends when $\epsilon =-\dot{H} /H^2= 1$, we deduce 
from Eq. (\ref{epsilonbeta}) that  
\begin{equation}
\label{betaend}
\beta_{end}=\frac{1-y_{end}^2}{1+y_{end}^2}   
\end{equation}
and $y_{end}$ is therefore maximized by minimizing the value of $\beta_{end}$. 

Similarly, Eq. (\ref{yendratio}) implies that the value of $y_{end}$ 
is maximized relative to its value $N$ e--foldings before the end 
of inflation by {\em minimizing} the area under the curve $\sqrt{\beta 
(\phi )}$ in the range $\phi = (\phi_N , \phi_{end} )$. 
In the low--energy limit, $\beta \rightarrow \epsilon_H$, and the 
conventional scenario is recovered. In this case, Liddle 
noted that the area under the curve $1/\sqrt{\epsilon_H (\phi )}$
is fixed by Eq. (\ref{yefolds}) once the number of e--foldings
is determined, since inflation ends precisely when $\epsilon_H =1$. 
It then follows that the smallest decrease in the Hubble parameter 
over the last $N$ e--foldings is achieved when $\epsilon_H$ 
is kept as small as possible. A similar conclusion holds 
for RSII inflation, although now it is $\beta$ that must be kept 
as small as possible. 

To proceed further, it is necessary to specify the behaviour of 
$\beta (\phi)$ during the final stages of inflation. 
We consider two cases in what follows, 
generalizing the method of Ref. \cite{liddle} to the braneworld scenario. 

\subsection{Case A: $\beta (\phi)$ increases monotonically} 

Liddle has argued convincingly that in the conventional scenario, $\epsilon$ 
should increase monotonically as the end of inflation approaches 
for a wide class of physically well--motivated models. 
This is also a very reasonable assumption to make in the 
braneworld scenario \footnote{Indeed, within the context of the slow--roll 
approximation, it follows that 
if the logarithmic derivative of the potential increases as the field 
rolls down its potential (as is to be expected for a single field 
inflation model that is undergoing a smooth exit), the slow--roll parameter 
(\ref{epsilonb}) also increases monotonically with decreasing $V(\phi)$.} 
It is straightforward to verify that 
the quantity $(1-y^2)/(1+y^2)$ also increases monotonically 
with decreasing $y$ and so it follows 
from Eq. (\ref{epsilonbeta}) that 
$\beta$ increases monotonically
if $\epsilon$ increases monotonically. 
Thus, the maximal 
value of $y_{end}$ follows in the limit where $\beta$ remains 
constant during the last $N$ e--foldings of inflation. 

When $\beta$ is constant, 
the integral in Eq. (\ref{yendratio}) may be readily evaluated, yielding
\begin{equation}
\label{maxint}
y_{end}^{max} = y_{N} \exp \left[ 
-\sqrt{\frac{4\pi \beta_N}{m_{4}^{2}}} (\phi_{end}-\phi_{N})\right]
\end{equation}
and 
integrating Eq. (\ref{yefolds}) and 
substituting the result into Eq. (\ref{maxint}) 
yields the maximum value for $y_{end}$ relative to $y_N$: 
\begin{equation}
\label{yendmaxA}
y_{end}^{max} = y_{N} \exp ( -N \beta_{N} )  .
\end{equation}
The maximum value for the Hubble parameter then follows
from Eq. (\ref{yHubble}):
\begin{equation}
\label{hmaxA}
H_{end}^{max} = \left(\frac{16 \pi\lambda}{3 m_{4}^{2}}\right)^{1/2}\frac{ 
y_{N} e^{ -N \beta_{N} }}{1- y_{N}^{2} e^{ -2N \beta_{N} }}   .
\end{equation}

It should be emphasized that the derivation of Eq. (\ref{hmaxA})
has not invoked the slow--roll approximation. 
In order to evaluate the upper limit, however, the parameters 
$\{ y_{N}, \beta_N , \lambda \}$ must be known. 
The direct dependence on the tension of the brane may be 
eliminated by expressing the upper limit (\ref{hmaxA}) 
in terms of the value of the Hubble parameter at $N$ e--foldings.  
The ratio is given by 
\begin{equation}
\label{pureHratio}
\frac{H_{end}^{max}}{H_{N}} = 
\frac{y_{end}^{max}}{y_{N}}\left(\frac{1-y_{N}^2}{1-\left( 
y_{end}^{max}\right)^2}\right)  
\end{equation}
and substituting Eq. (\ref{yendmaxA}) 
results in the upper limit on the energy scale 
at the end of inflation: 
\begin{equation}
\label{Hratio}
\frac{H_{end}^{max}}{H_{N}} =  e^{ -N \beta_{N} }\frac{ 1- y_{N}^2}{1- 
y_{N}^{2} e^{ -2N \beta_{N} }}  .
\end{equation}
In principle,   
$\beta_{N}$ and $y_{N}$ can be determined 
in terms of the potential and its derivatives. 

\subsection{\bf Case B: $\beta (\phi )$ and $\beta'(\phi )$ increase 
monotonically}

The second case assumes that both $\beta (\phi )$ and 
$\beta' (\phi)$ increase monotonically \cite{liddle}. This 
results in a potentially stronger constraint than that of case A since the 
area under the curve $\sqrt{\beta (\phi)}$ in 
Eq. (\ref{yendratio}) is enhanced. Given these restrictions,
the form of $\beta$ that
maximizes the number of e--folds and 
the energy scale is linear:
\begin{equation}
\label{betaphi}
\beta= B \phi  ,
\end{equation}
where, without loss of generality, 
we may assume that $B$ is a positive constant and $\phi >0$. 
This implies that $\beta' >0$ and 
hence $\beta > \gamma$. The 
total number of e--foldings consistent with 
this behaviour follows from Eqs. (\ref{betagamma}) and (\ref{yefolds}): 
\begin{equation}
\label{Nupper}
N = \frac{\sqrt{\beta_{end}}-\sqrt{\beta}}{\sqrt{\beta} 
(\beta -\gamma)}
\end{equation}
and this implies that $\beta$ must be sufficiently 
close to $\gamma$ to ensure that enough inflation occurred. 

An expression for $\phi_{N}$ can be found by rearranging 
Eq. (\ref{betagamma}) and noting that 
$B= \beta_{end}/\phi_{end}$:
\begin{equation}
\label{phiN}
\phi_{N}^{2} = \frac{m_{4}^{2}}{16\pi}\frac{\beta_N}{(\beta_N-\gamma_N)^2}  .
\end{equation}
Likewise, an expression for $\phi_{end}$ 
follows from 
$\phi_{end} =  \beta_{end}\phi_{N}/\beta_{N}$:
\begin{equation}
\label{phiend}
\phi_{end}=\sqrt{\frac{m_{4}^{2}}{16\pi}}\frac{1}{\sqrt{\beta_N}}
\frac{\beta_{end}}{(\beta_N-\gamma_N)}  .
\end{equation}
Given Eq. (\ref{betaphi}), 
the integral in Eq. (\ref{yefolds}) can be evaluated and rearranged to yield
\begin{equation}
\label{Ncubedbit}
\frac{N(\beta_N-\gamma_N)+1}{\sqrt{\beta_{end}}}=\frac{1}{\sqrt{\beta_N}} 
\end{equation}
and the integral in Eq. (\ref{yendratio})
reduces to 
\begin{equation}
\label{intreduce}
\frac{y_{end}}{y_{N}} = \exp\left[-\sqrt{\frac{16\pi B}{9m_{4}^2}}
\left( \phi_{end}^{3/2} - 
\phi_{N}^{3/2}\right)\right] .
\end{equation}
Substituting Eqs. (\ref{phiN}) and (\ref{phiend}) into Eq. (\ref{intreduce}) 
implies that  
\begin{equation}
\label{int1}
\frac{y_{end}}{y_{N}} = \exp \left[ -
\frac{1}{3}\frac{\beta_N}{\beta_N-\gamma_N}
\left( 
\frac{\beta_{end}^{3/2}}{\beta_N^{3/2}}-1 \right) \right] 
\end{equation}
and it follows, after substituting 
Eq. (\ref{Ncubedbit}) into Eq. (\ref{int1}), that
\begin{eqnarray}
\label{yendmaxB}
y_{end}^{max} = y_{N}\exp \left[ 
- \frac{\beta_{N}}{3(\beta_{N}-\gamma_{N})}
\left( \left[ 1 + \right. \right. \right. \nonumber \\
\left. \left. \left. 
 N(\beta_{N}-\gamma_{N}) \right]^{3}-1\right) \right]  .
\end{eqnarray}
Eq. (\ref{yendmaxB}) can be used to find the
ratio of the 
Hubble parameters
$H_{end}^{max}$ and $H_{N}$, as shown 
in Eq. (\ref{pureHratio}), by substituting 
Eq. (\ref{yendmaxB}) where required. Case 
A can be recovered from Case B in the limit that $\beta_{N} \rightarrow 
\gamma_{N}$, i.e., $\beta' \rightarrow 0$.

A particular form for the function $y(\phi)$ 
motivated by particle physics considerations could 
now be chosen and an estimate for the 
energy scale at the end of inflation deduced. However, 
given the enhanced
accuracy of recent CMB measurements, 
it is of more interest to consider what observations 
may reveal about the underlying inflationary model. 
We therefore explore this possibility in the following Section.

\section{Observational Constraints on the End of Inflation}

In the previous Section upper limits on the energy scale 
at the end of RSII inflation were derived without limiting the analysis 
to the slow--roll regime. However, 
it is necessary to assume that the slow--roll 
approximation applied during the epoch when observable scales went 
beyond the Hubble radius, 
since 
the derivations of the perturbation spectra are only valid 
in this regime.  
In the slow--roll limit, Eq. (\ref{ydefine}) 
reduces to
\begin{equation}
\label{yofv}
y^2 =\frac{V}{V +2\lambda} 
\end{equation}
and Eqs. (\ref{epsilonb}) and (\ref{epsilonbeta}) then 
imply that 
\begin{equation}
\label{ebsro}
\epsilon_B = \left( 1+ \frac{V}{\lambda} \right) \beta  .
\end{equation}

The amplitudes of the 
scalar and tensor perturbations are given by \cite{maartens,lmw}
\begin{eqnarray}
\label{deltaS}
A_S^2 = \frac{1}{25\pi^2} \frac{H^4}{\dot{\phi^2}} 
\\
\label{deltaT}
A^2_T =\frac{4}{25\pi m^2_4} H^2F^2 ,
\end{eqnarray}
respectively, where 
\begin{equation}
\label{Fdefine}
\frac{1}{F^2} = \sqrt{1+s^2} -s^2 {\rm sinh}^{-1} 
\left( \frac{1}{s} \right)
\end{equation}
and 
\begin{equation}
\label{sdefine} 
s \equiv \left( \frac{3H^2m_4^2}{4\pi \lambda} \right)^{1/2}  .
\end{equation}
The function (\ref{Fdefine}) 
may be viewed as a function of the inflaton potential. 

The corresponding spectral indices (tilts) 
are given by \cite{maartens,hueylid1}
\begin{eqnarray}
\label{ns}
n_S -1 = -6\epsilon_B +2\eta_B  \\
\label{r}
r = -8 n_{T}  ,
\end{eqnarray}
where $r \equiv 16A^2_T/A_S^2$ \footnote{We employ 
the normalization conventions of Ref. \cite{infrev1}. The ratio of the 
tensor to scalar amplitudes is related to the quantity, $r$, employed 
by Peiris {\em et al.} \cite{wmap1} by $r=16A^2_T/A^2_S$.}. 
Eq. (\ref{r}) represents the consistency equation for RSII 
inflation \cite{hueylid1}. 

It follows from Eqs. (\ref{betadefn}), (\ref{deltaS}) and 
(\ref{deltaT}) that the ratio of tensor to scalar amplitudes
can be expressed directly in terms of 
the first braneworld flow parameter: 
\begin{equation}
\label{betar}
\beta = \frac{r}{16F^2}   .
\end{equation} 

We now
consider in turn the two cases discussed in 
Section 3.  

\subsection{Case A}

Substituting Eqs. (\ref{yofv}) and (\ref{betar})
into Eq. (\ref{Hratio}) allows the upper limit 
on the inflationary energy scale 
for case A to be expressed in terms of 
the relative amplitude of the gravitational wave perturbations
\begin{equation}
\label{caseA}
\left. 
\frac{H^{max}_{end}}{H_N} = 
\frac{\exp \left( - \frac{rN}{16F^2} \right)}{1+x \left[ 
1- \exp \left( - \frac{rN}{8F^2} \right) \right]} \right|_N
\end{equation}
where $x \equiv V/(2\lambda )$ 
and quantities on the right hand side of Eq. (\ref{caseA}) 
are to be evaluated $N$ e--foldings before 
the end of inflation. 

Thus far, only the slow--roll approximation has been invoked
and  the magnitude of the inflaton 
potential relative to that of the brane tension
has not been specified.
We now focus on
the high--energy limit $(V \gg \lambda , x \gg 1, s\gg 1)$, corresponding to 
the region of parameter space relevant to the 
steep inflationary scenario \cite{steep}. 
In this limit, the function (\ref{Fdefine}) 
simplifies 
considerably, 
$F^2 \approx [27H^2m_4^2/(16\pi \lambda )]^{1/2} \approx 3x \approx 3s/2$ 
and Eq. (\ref{caseA}) reduces to 
\begin{equation}
\label{hma}
\frac{H^{max}_{end}}{H_{N}}=\frac{e^{-m_A/x}}{1+x\left(1- e^{-2m_A/x}\right)},
\end{equation} 
where 
\begin{equation}
\label{madef}
m_A \equiv \frac{rN}{48}  .
\end{equation}

In the high energy limit, the 
gravitational wave amplitude varies as 
$A_T^2 \propto 
(\lambda /m_4^4) (V/2\lambda)^3$. Thus, 
a positive detection of the tensor perturbations would not 
by itself constrain the magnitude of the inflaton potential 
unless the brane tension has also been specified. However, 
at present the tension is viewed as a free parameter
since a definitive particle physics model for inflation has yet to be 
developed.  
In view of this, and anticipating a 
future detection of the gravitational wave spectrum, we 
consider how the bound (\ref{hma}) varies with 
$x=V/(2 \lambda)$ 
for given values of $\{ r , N \}$. 

Since $r$ is small, the high--energy 
limit of Eq. (\ref{hma}) may be approximated 
by expanding the exponential terms to first--order 
in a Taylor series:
\begin{equation}
\label{caseAlimit}
\frac{H^{max}_{end}}{H_N} = \frac{1-rN/(48x_N)}{1+rN/24}  .
\end{equation}
We therefore arrive at the remarkably simple result that 
the maximal energy scale at the end of steep inflation can not 
exceed 
\begin{equation}
\label{asymptotic}
\frac{H^{max}_{end}}{H_N} = \frac{1}{1+rN/24}  .
\end{equation}
Thus, at sufficiently high energy scales, the upper 
limit is determined entirely by the relative gravitational 
wave amplitude for a given number of e--foldings. Increasing the gravitational 
wave amplitude strengthens the bound
on the energy scale at the end of inflation. 
We consider this limit in more detail in Section 4.3.

\subsection{Case B}

For this case, both braneworld flow parameters (\ref{betadefn}) 
and (\ref{gammadefn}) 
need to be considered. To arrive at 
a bound given in terms of observable parameters,  
we restrict the analysis to the high energy limit 
$(V \gg 2\lambda, x\gg 1 )$ and 
express Eq. (\ref{yendmaxB}) in terms of the inflaton
potential and its derivatives. 
These quantities can then be related to the spectral index of the 
scalar perturbation spectrum and the gravitational wave amplitude. 
The scalar spectral index is deduced 
from Eqs. (\ref{epsilonb}), (\ref{etab}) and (\ref{ns}):
\begin{equation}
\label{highens}
n_{S}-1 =  -\frac{\lambda m_{4}^{2}}{2\pi V}\left[3 \frac{V'^2}{V^2} - 
\frac{V''}{V}\right] 
\end{equation}
and the gravitational wave amplitude follows 
from Eq. (\ref{betar}):
\begin{equation}
\label{betav}
\beta = \frac{m_4^2}{4\pi} \frac{\lambda^2V'^2}{V^4}
= \frac{r\lambda}{24V} .
\end{equation}

The relationship (\ref{betagamma}) between the braneworld 
flow parameters implies that the combination
$(\beta-\gamma )$ can be expressed 
in terms of $\beta$ and its first derivative $\beta'$. Hence, by 
employing Eq. (\ref{betav}), the bound (\ref{yendmaxB}) can be written as
\begin{eqnarray} 
\label{yendmaxBV}
y_{end}^{max} = y_{N}\exp\left[ 
\frac{\frac{2\lambda^2}{\kappa^2}\frac{V'^2}{V^4}}{3\left( 
\frac{2\lambda}{\kappa^2}\frac{V''}{V^2} - 
\frac{4\lambda}{\kappa^2}\frac{V'^2}{V^3}\right) } 
\left(\left[1 - \right. \right. \right. \nonumber \\
\left. \left. \left. 
N\left(\frac{2\lambda}{\kappa^2}\frac{V''}{V^2} - 
\frac{4\lambda}{\kappa^2}\frac{V'^2}{V^3}\right)\right]^3-1\right)\right]_N
\end{eqnarray}
where $\kappa^2 \equiv 8\pi/m_4^2$. 
Eq. (\ref{yendmaxBV}) can then be expressed succinctly in terms 
of the scalar spectral index and gravitational wave amplitude
by substituting 
Eqs. (\ref{highens}) and (\ref{betav}):
\begin{equation}
y_{end}^{max} = y_{N} e^{-m_B/x_N}  ,
\end{equation}
where the constant 
\begin{eqnarray}
\label{mbdef}
m_B  \equiv - \frac{r}{72(n_S-1) +6r}
\left[ \left(  1- \right. \right. \nonumber \\
\left. \left. \frac{N}{2}\left( n_S-1+\frac{r}{12} 
\right)\right)^3 
-1 \right]  
\end{eqnarray}
is determined by $n_S$ and $r$. We therefore deduce that 
\begin{equation}
\label{HratioslowrollB}
\frac{H_{end}^{max}}{H_{N}}=\frac{e^{-m_B/x_N}}{1+x_N
\left(1- e^{-2m_B/x_N}\right)}.
\end{equation}

It should be emphasized that the bound 
(\ref{yendmaxB}) follows from the assumption that the braneworld 
flow parameter (\ref{betadefn}) and its first derivative 
are monotonically increasing functions. In this case,  Eq. (\ref{betagamma}) 
implies that $\beta > \gamma$ and this is 
equivalent to imposing the restriction 
$1-n_S > r/12$ at the level of approximation 
considered in this Subsection. The bound (\ref{HratioslowrollB})
is valid only in this region of parameter space. 

To summarize, the bounds (\ref{hma}) and 
(\ref{HratioslowrollB}) on the steep inflationary energy scale 
may be written in a unified form, 
where each limit is determined by the numerical value of the parameters 
$m_{A,B}$. At sufficiently high energy scales $(x\gg 1)$, 
the relative bounds asymptote to the 
constant values: 
\begin{equation}
\label{maxAB}
\frac{H_{end}^{max}}{H_{N}} = \frac{1}{1+2m_i}  ,
\end{equation}
where $i = \{ A,B \}$. 

\subsection{Parameter Estimates}

\begin{figure}
\includegraphics[width=8.5cm]{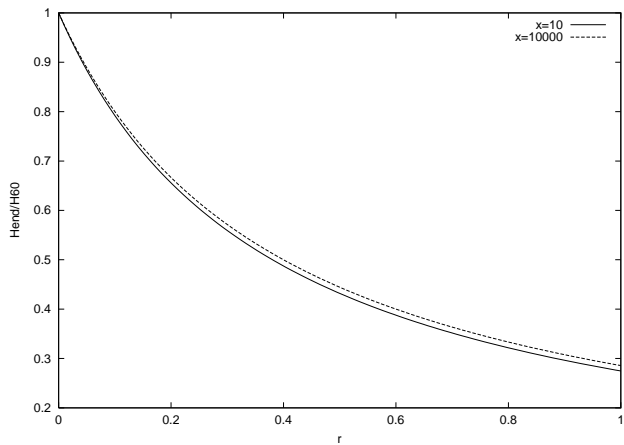}%
\caption{Illustrating Eq. (\ref{caseA})
for case A studied in the text. The  
maximal value of the Hubble parameter 
at the end of braneworld inflation, $H^{max}_{end}/H_N$, 
is shown on the vertical axis and   
the gravitational wave amplitude, $r$, on the horizontal axis.   
The bound is parametrized
relative to the value of the Hubble parameter when 
observable scales first crossed the Hubble radius during inflation and 
this is assumed to have occurred 60 e--foldings before the 
end of inflation. 
The solid line corresponds to $V_N/2\lambda=10$ and the dashed 
line to 
$V_N/2\lambda=10^4$. There is 
little difference despite the three orders of
magnitude increase in $V_N/2\lambda$.}  
\label{figure1}
\end{figure}

\begin{figure}
\includegraphics[width=8.5cm]{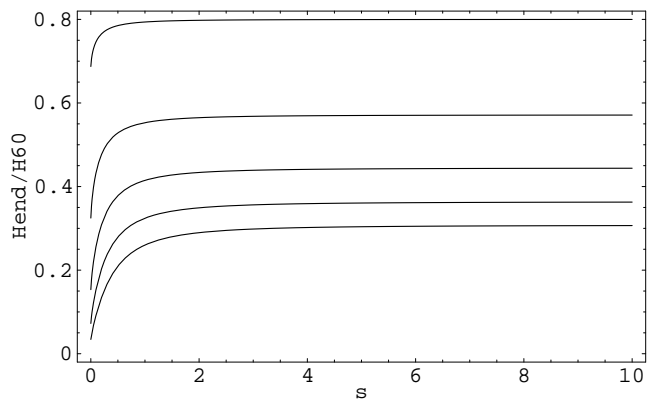}%
\caption{Illustrating the dependence of the 
relative bound (\ref{caseA}) 
on the energy scale  
for fixed 
gravitational wave amplitudes: 
$r =0.1$ (top plot), $r=0.3$, $r=0.5$, $r=0.7$ and $r=0.9$ (bottom plot).
The variable $s$ is defined in Eq. (\ref{sdefine}).
Qualitatively similar behaviour arises for case B. 
In both cases, the ratio $H_{end}^{max}/H_{60}$ 
asymptotes to a constant value that is determined by the magnitude of 
the tensor perturbations.}  
\label{figure2}
\end{figure}

\begin{figure}
\includegraphics[width=8.5cm]{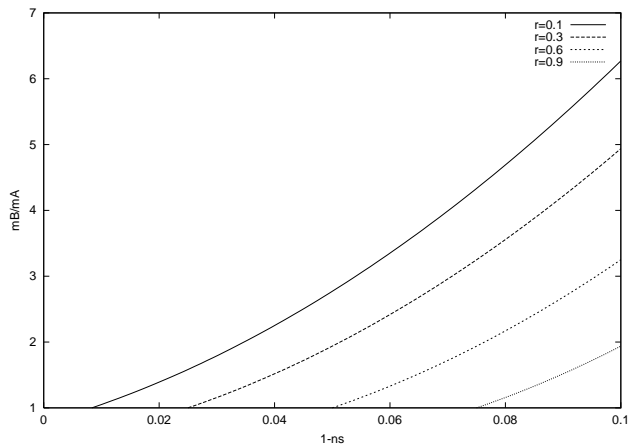}%
\caption{Illustrating the relative values of the 
parameters $m_A$ and $m_B$ defined in the text in Eqs. (\ref{madef}) 
and (\ref{mbdef}) for different spectral tilts and gravitational 
wave amplitudes.}  
\label{figure3}
\end{figure}

It is necessary to consider a couple of provisos
when interpreting the data from recent measurements of the CMB to estimate the
energy scale from the constraints derived above. 
In general, the metric perturbations on large scales  
are not necessarily held constant in RSII cosmology 
due to their interaction with
the bulk gravitational field \cite{bulk1,bulk2}. This induces a 
backreaction that is manifested as 
a non-local source of energy--momentum on the brane and alters the 
dynamics.
The possible effects of the backreaction on the CMB anisotropies is also a 
factor that may result in  differences from the
standard interpretation of the observable parameters \cite{backreaction}.  
However, it is consistent to assume the backreaction 
is negligible at linear order \cite{deruelle} and we
take the 
parameter estimates from WMAP as our starting point \cite{wmap1}. 
 
To arrive at numerical estimates for the bounds, we 
must specify the number of e--foldings, $N$, 
that corresponds to observable 
scales. There is some uncertainty in this value given its dependence 
on the reheating temperature. Recently, upper limits on $N$ were 
estimated for standard inflation, where it was concluded that 
a reliable range is $50 < N <60$ in the absence of 
a specific inflationary model \cite{dod,lid}. In the RSII braneworld 
scenario, this may alter slightly if inflation ends in the 
high--energy regime, since the evolution of the scale factor 
immediately after 
inflation is different. Here we specify $N=60$. 
For case
A, the asymptotic value (\ref{asymptotic}) is only increased 
by a factor of $(12+30r)/(12+25r)$ 
by choosing  $N=50$. This relaxes the upper 
bound by $3 \%$ for $r=0.1$ and 
by 13$\%$ for $r=0.9$. 

The upper bound for case A is stronger for higher gravitational 
wave amplitudes. We therefore take the weakest upper limit of $r \le 0.9$
quoted by WMAP from the combined data set \cite{wmap1}. 
Fig. \ref{figure1} illustrates the bound as a function of 
gravitational wave amplitude for two different energy scales. 
The bound is fairly insensitive 
to the energy scale corresponding to observable scales. 
Fig. \ref{figure2} illustrates the bound with varying 
energy scale for fixed values of $r$ in the range 
$r \in [0.1, 0.9]$. When $r=0.9$, 
the energy scale at the end of inflation
is constrained to be no more than 30$\%$ that of the observable 
scale.

A direct comparison between cases A and B in the high energy limit 
is made possible by comparing the relative magnitude of 
the asymptotic values (\ref{maxAB}) for given gravitational 
wave amplitudes. It follows from Eqs. (\ref{madef}) and (\ref{mbdef}) 
that 
\begin{eqnarray}
\label{mratio}
\frac{m_B}{m_A} =- \frac{8}{N[ 12(n_S-1)+r]} \left[ \left( 
1- \right. \right. \nonumber \\
\left. \left. 
\frac{N}{2} \left( n_S -1 +\frac{r}{12} \right) \right)^3 -1 \right]
\end{eqnarray}
and Fig. \ref{figure3} illustrates the dependence of this 
ratio on observable parameters. 
The two cases become degenerate, $m_B \rightarrow 
m_A$,  in the limit 
$r \rightarrow 12(1-n_S)$. The bound on the energy 
scale for case B is made stronger by 
increasing the difference between $r$ and $(1-n_S)$ and 
the ratio between the two bounds is enhanced for smaller gravitational 
wave amplitudes. Moreover, the ratio increases as 
the scalar perturbation spectrum is tilted away from 
a Harrison--Zeldovich form. For example, taking $n_S =0.94$ as the smallest 
value consistent with WMAP \cite{wmap1} implies that 
$m_B/m_A = 3.4$ for $r=0.1$ and $m_B/m_A = 2.2$ for $r=0.35$. 
This enhances the upper limit on the 
energy scale by 68$\%$ and 64$\%$, respectively.  

\section{Conclusion}

The energy scale at which inflation ends is not precisely known.
Until the correct particle physics theory is
identified and this scale determined, 
our understanding of the dynamics of the very 
early universe is incomplete. It is 
therefore important to constrain the energy scale at 
at the end of inflation. 

By introducing 
new `braneworld flow parameters' that generalize the Hubble flow parameters 
of the standard scenario, 
we have derived an upper bound on the energy scale at the end of  
RSII inflation 
by invoking only very weak assumptions about the form of the 
inflaton potential, namely that the first braneworld flow parameter, 
$\beta (\phi )$, is a monotonically increasing function during the final 
stages of inflation. This should be the case for 
a wide class of realistic potentials which are 
smooth, differentiable, monotonically decreasing functions. 
By further assuming that observable scales 
correspond to the 
high energy and slow--roll regimes relevant to the steep inflationary 
scenario \cite{steep}, this
upper bound was expressed in terms of observable parameters. 
A positive detection of the gravitational 
wave background would determine the energy scale 
when observable modes crossed the Hubble radius.
In the absence of such a detection, 
we have expressed the bounds on the scale at the 
end of inflation relative to this scale. 

At progressively higher energy scales, this relative 
bound
is 
only very weakly dependent 
on the magnitude of the inflaton potential
and asymptotes to a constant value. 
The asymptotic limit 
depends on the gravitational wave amplitude and the constraint 
becomes tighter as $r$ increases. This is qualitatively 
similar behaviour to the standard scenario \cite{liddle}. 
The amplitude of the scalar perturbations varies as $A_S^2 \propto 
H^2_N/\beta_N$ and increasing $\beta_N$ (and consequently $r$) 
for a given COBE normalization increases 
the energy scale $N$ e--foldings before the end of inflation. 
On the other hand, it follows from Eq. (\ref{yendmaxA}) 
that this has the effect of 
reducing the maximal energy scale at the end of inflation. Hence, 
maximizing $H_N$ does not maximize $H_{end}$. 

This has implications for models of steep inflation,
where the logarithmic slope of the potential is large. 
Whilst this may be viewed as a positive feature, in that it leads to 
an amplitude of gravitational waves that is potentially
observable \cite{steep}, it results in a stronger upper 
limit on $H_{end}$. Consequently, when reheating proceeds through 
gravitational 
particle production, the era immediately after inflation 
where the universe 
is dominated by the kinetic energy of the inflaton field is prolonged.  
This makes is harder to satisfy the bounds imposed by primordial 
nucleosynthesis
on excessive gravitational wave contributions to the 
energy density \cite{sami,kostas}. 

In this paper, 
we have considered the region of parameter space where observable 
scales correspond to the high-energy regime, i.e., where the 
quadratic contribution in the Friedmann equation (\ref{Friedmann}) 
dominates the dynamics. 
However, it is possible that for some models, 
the last $N$ e--foldings of inflation occur when 
this term is becoming sub--dominant. In the low--energy regime 
($\rho/2\lambda \ll 1$), the limits of the standard scenario are recovered
\cite{liddle}, since the braneworld parameters 
$\{ \beta , \gamma \} \rightarrow \{ \epsilon_H , \eta_H \}$. 
It should be emphasized, however, that the bounds 
derived in Section 3, Eqs. (\ref{Hratio}) and (\ref{yendmaxB}), 
are valid over all energy scales.  
Moreover, it follows from the definitions (\ref{betadefn}) and 
(\ref{gammadefn}) that the relation  
\begin{equation}
\label{interesting}
\gamma -\beta = \eta_B - 2\epsilon_B
\end{equation}
is also valid for all energy regimes when the field is 
slowly rolling. Since the 
second flow parameter (\ref{gammadefn}) 
only appears in the 
bound (\ref{yendmaxB}) through the combination $(\gamma -\beta )$, 
it follows that 
Eqs. (\ref{ebsro}) 
and (\ref{interesting}) are sufficient 
to relate the braneworld flow parameters 
directly to the corresponding braneworld slow--roll parameters, 
(\ref{epsilonb}) and (\ref{etab}),
and hence the scalar spectral index and gravitational wave amplitude. 
In principle, therefore, the bound for case B derived in Section 4.2 could 
be extended to cover all energy scales.

Finally, the upper bounds on the energy scale for RSII inflation
followed once a variable was identified 
that played the role of the Hubble flow parameter (\ref{epsilon}). 
In principle, therefore, other scenarios where the Friedmann 
equation is modified may be developed along similar lines
to those of the present
work.  
One extension of the RSII scenario is to relax the 
assumption of a reflection symmetry in the bulk dimension. 
This introduces a further term in the Friedmann equation 
that scales as $\rho^{-2}$ \cite{othermoves,asy}. 
The inclusion of a Gauss--Bonnet combination of curvature 
invariants in the bulk action also modifies the dynamics and  
at high energies the Friedmann equation 
asymptotically takes the form $H^2 \propto 
\rho^{2/3}$ \cite{gbf}. 
The implications of this modification for inflation
were recently considered \cite{ln}. It would be interesting to 
develop a framework involving generalized flow parameters 
for these scenarios and to investigate how these new 
features alter the bounds on the inflationary energy scale. 

\vspace{1cm}

\begin{acknowledgments}
RMH is supported by the Particle Physics 
and Astronomy Research Council (PPARC).
JEL is supported by the Royal Society. 

\end{acknowledgments}

\end{document}